\documentclass[12pt]{iopart}

\expandafter\let\csname equation*\endcsname\relax
\expandafter\let\csname endequation*\endcsname\relax

\usepackage{amsmath}
\usepackage{amssymb, bm, graphicx}
\usepackage{color}

\begin{document}
\title[Self-similar rupture of thin films of power-law fluid]{Self-similar rupture of thin films of power-law fluid}
\author{Michael C. Dallaston, Steven A. Kedda, and Scott W. McCue}
\address{School of Mathematical Sciences, Queensland University of Technology, Brisbane, QLD, 4001, Australia.}
\ead{michael.dallaston@qut.edu.au}
\vspace{10pt}
\begin{indented}
\item[]\today
\end{indented}

\begin{abstract}
Models that describe Newtonian liquid films evolving due to the competing effects of surface tension and attractive intermolecular or van der Waals forces are known to rupture in finite time in a self-similar manner.  We extend the computation of similarity solutions to non-Newtonian power-law liquid films.  The resulting bifurcation diagram, indexed by power law exponent $n$, has a highly nontrivial structure with branches merging via a snaking bifurcation around $n=1$.  A countably infinite number of solutions are also found in the extreme shear-thinning ($n\to 0$) limit, in which similarity solutions possess an exponentially small inner region.  Numerical simulations of the time-dependent model are shown to be attracted to the single primary branch of similarity solutions.   The asymptotic structure of solutions in the small $n$ limit is also determined.

\end{abstract}

\section{Introduction}

The study of the partial differential equation (PDE) models that arise from thin liquid film dynamics is extensive \cite{bonn2009wetting,craster2009dynamics,oron1997long-scale}.  Models that use lubrication theory to simplify the full Navier--Stokes equations and boundary conditions are readily modified to include any number of physical effects, such as gravity \cite{huppert1982propagation,huppert2006gravity}, surface tension \cite{myers1998thin}, thermo- or solutal Marangoni forces \cite{gaver1990dynamics,jensen1992insoluble,kedda2024long,oron1997long-scale,shklyaev2010superexponential,shklyaev2012long}, or intermolecular disjoining pressures~\cite{erneux1993nonlinear,ruckenstein2018spontaneous,witelski1999stability,zhang1999similarity,ida1996thin}.  Even for Newtonian fluids (which forms the majority of the literature), the resulting equations are highly nonlinear, fourth-order in space (when surface tension is included), and can exhibit many unexpected properties related to finite-time singularities \cite{dallaston2021regular,dallaston2018discrete}.

For very thin films, gravity may be neglected, and instead the intermolecular or van der Waals forces that exist between a fluid and the solid wetted substrate become important.  It is these forces that will drive a thin film of fluid to rupture (dewet) and subsequently create dry spots.  It is standard in lubrication theory to integrate the body forces over the film thickness to derive an equivalent disjoining pressure acting on the interface \cite{teletzke1987liquids}.  Including both attractive and repulsive effects in the disjoining pressure provides a way to model the creation of a precursor film and retraction of a contact line post-rupture \cite{becker2003complex,ghatak1999dynamics,glasner2003coarsening}.  However, the dynamics leading up to rupture will be dominated by the attractive force.

A spatially fourth-order PDE that includes the competition between surface tension and disjoining pressure for a Newtonian fluid film was considered in \cite{witelski1999stability,zhang1999similarity}, who analysed the finite-time rupture of the film in the context of self-similarity; as the rupture time and location are approached, the film thickness tends to a profile that is constant in shape under an appropriate time-dependent scaling.  While the exponents of the scaling can be found analytically by balancing the time dependence of each term (a first-kind self-similarity \cite{barenblatt1996scaling,eggers2015singularities}), similarity profiles can only be calculated numerically as the solution of a nonlinear boundary-value problem.   It turns out there are countably infinite similarity profiles, only one of which is linearly stable to perturbations, and thus the attractor for PDE simulations, under a suitable definition of stability~\cite{witelski1999stability}.  Further extensions of this problem include the application of numerical continuation to construct multiple similarity solutions \cite{tseluiko2013homotopy}, the generalisation of disjoining pressure terms to other exponents \cite{dallaston2021regular,dallaston2018discrete,dallaston2017self}, exponential asymptotic selection problem for large branch number \cite{chapman2013exponential}, and the asymptotic analysis of branch merging in the generalised problem \cite{chapman2023role}.

Many industrial and biological fluids involved in lubrication flows exhibit non-Newtonian behaviour \cite{eley2005applied,myers2005application,zhang2003analysis}.  One commonly used non-Newtonian model is the power-law model, in which the stress is a power-law function of the strain rate, with the power $n$ a fluid-dependent parameter characterising the nature of the non-Newtonian response ($n <1$, $n>1$, and $n=1$ giving shear-thinning, shear-thickening, and Newtonian behaviour, respectively).  The evolution of non-Newtonian fluids of power-law type due to surface tension has been considered in the context of contact-line motion~\cite{flitton2004surface,king2001two}, as well as capillary levelling of an almost-flat interface~\cite{dallaston2024capillary,iyer1996leveling}.  There are also a number of results concerning flow down an inclined surface, in which gravity and surface tension are both included \cite{miladinova2004thin,noble2013thin}.

In this article we focus on rupture of thin films of power-law fluids, bounded between a solid substrate and free surface, due to van der Waals forces.  This problem has been numerically simulated in \cite{hwang1993rupture} for a plane substrate, and in \cite{gorla2001rupture} for a cylindrical substrate.  The model of a tear film in \cite{zhang2003analysis} includes both power-law fluids and van der Waals forces. In each of these studies, finite-time rupture is observed, but solutions are not analysed for self-similar behaviour in particular.  In addition to numerical simulations, \cite{garg2017self-similar} determine appropriate similarity exponents, but do not explicitly compute similarity profiles.  There are thus still significant open questions regarding the calculation and properties of similarity solutions describing rupture, particularly as dependent on the power-law exponent $n$.

 Apart from coating of solid substrates, rupture of free thin liquid sheets of power-law fluids (featuring two interfaces, and no solid substrate) have also been considered~\cite{thete2015self,garg2022local}.  The dominant physical effects, and thus the self-similar scalings, are highly dependent on power-law exponent $n$.  When inertia is present \cite{thete2015self}, the dominant balance is either (for $n> 6/7$) between inertia, viscosity, and van der Waals force or (for $n < 6/7$) between inertia, surface tension, and van der Waals force.  When inertia is neglected \cite{garg2022local}, the dominant forces are either viscosity and van der Waals force (for $n \gtrsim 0.58$) or viscosity, surface tension, and van der Waals force (for $n \lesssim 0.58$).  Similarity solutions (which are of the second kind) have only been explicitly computed in the case where the dominant terms are the viscous and van der Waals terms.

In this article we explicitly compute similarity solutions for the rupture of power-law fluids on a substrate due to van der Waals forces.  By starting at the countable similarity solutions that exist for Newtonian fluids~\cite{witelski1999stability,zhang1999similarity}, we construct a bifurcation diagram of branches of similarity solutions in the power-law exponent $n$.  This bifurcation diagram exhibits a highly nontrivial structure, with all but a finite number of branches merging in the neighbourhood of $n=1$ as $n$ is increased or decreased.  A single primary branch is shown to be the attractor by comparing the results of numerical simulations of the evolution equation with similarity solutions.  Only the primary branch continues to exist in both the extreme shear-thickening ($n\to\infty$) and shear-thinning ($n \to 0$) limits.

We also consider the small-$n$ asymptotic limit in some detail.  Severe challenges in the numerical computation of similarity solutions are uncovered as $n$ becomes small.  By performing the asymptotic analysis in the limit $n\to 0$, we determine the presence of an exponentially small inner region.  This asymptotic structure is similar to one discovered in a model of ice stream surges that was described in \cite{fowler1996ice} and further analysed in \cite{fowler2000small,kember2000exponential,obrien1998asymptotics}; by contrast, our model is fourth order in space, and even the leading-order problem can only be solved numerically.  However, with this approach we are able to formulate and solve this leading-order problem for the similarity profiles as $n\to 0$, determining there are a countably infinite number of solutions in this limit.  This asymptotic analysis also motivates a rescaled version of the system that allows us to numerically compute solution branches for small but nonzero $n$, which is necessary to complete the bifurcation diagram.

In the next section we summarise the formulation of the power-law thin-film equation with disjoining pressure and surface tension.  In section \ref{sec:numerics} we describe our numerical method for solving this equation and present results that demonstrate the generic tendency to self-similar rupture.  In section \ref{sec:selfsim} we derive the boundary-value problem for similarity profiles, describe our numerical solution procedure, and construct the bifurcation diagram for this system as similarity solutions vary depending on the power-law exponent $n$.  In section \ref{sec:asymptotics} we derive the asymptotic structure of the problem in the $n\to 0$ limit, and numerically compute the countably-infinite number of leading-order solutions in this limit.   Concluding remarks are provided in section \ref{sec:discussion}.

\section{Formulation}
\label{sec:formulation}

\begin{figure}
\centering
\includegraphics{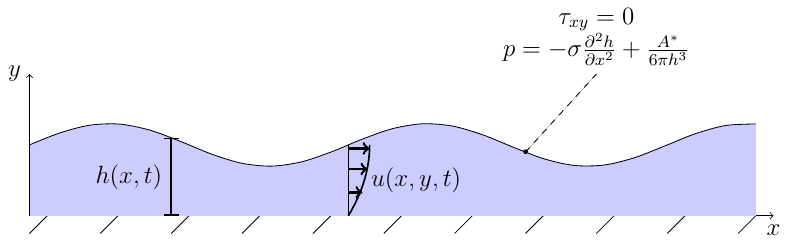}
\caption{A film of power-law fluid of thickness $h(x,t)$ wets a solid surface and flows due to surface tension and disjoining pressure applied to the fluid-air interface, while shear stress $\tau_{xy}$ vanishes at the interface.}
\label{fig:schematic}
\end{figure}

Here we summarise the formulation of the thin film equation with power-law rheology; for a more extensive derivation, see \cite{garg2017self-similar,kedda2025mathematical}.  Consider a solid surface completely wetted by a fluid of power-law rheology, with a film sufficiently thin that gravity is negligible, and intermolecular attraction between the fluid and solid is important.  We assume plane-symmetry, so that there is a single coordinate $x$ tangent to the solid surface, and a normal coordinate $y$.  Figure \ref{fig:schematic} depicts the system under consideration.  

Denoting the fluid-air interface as $y = h(x,t)$, conservation of mass implies that
\begin{equation}
\frac{\partial h}{\partial t} + \frac{\partial q}{\partial x} = 0,
\end{equation}
where the flux $q$ is the integral of the horizontal velocity component $u$:
\[
q = \int_0^h u \, \mathrm dy.
\]
The fluid flow is governed by the non-Newtonian version of the Navier--Stokes equations. Under the lubrication approximation, the pressure $p$ is independent of the vertical direction $y$ (in the absence of body forces) and the dominant balance in the horizontal momentum equation is between the pressure gradient and the shear stress:
\[
\qquad \frac{\partial p}{\partial x} = \frac{\partial \tau_{xy}}{\partial y}, \qquad 0 < y < h,
\]
where the viscous stress $\tau_{xy}$ is related to the strain rate according to the assumed rheology of the fluid.  Under the lubrication approximation, the dominant term in the strain rate is the vertical gradient of the horizontal velocity component.  In the case of a power-law fluid, the stress is therefore given by
\[
\tau_{xy} = M \left|\frac{\partial u}{\partial y}\right|^{n-1}\frac{\partial u}{\partial y}, \qquad 0 < y < h,
\]
where $M$ is the consistency coefficient and $n$ is the power-law exponent.  A Newtonian fluid corresponds to the special case $n=1$, in which case $M$ is the viscosity.

Applying the zero shear-stress condition ($\tau = 0$) to the interface $y=h$, 
\[
\tau = -\frac{\partial p}{\partial x}(h-y), \qquad \frac{\partial u}{\partial y} = -M^{1/n}\left| \frac{\partial p}{\partial x} \right|^{1/n-1}\frac{\partial p}{\partial x} (h-y)^{1/n}.
\]
Integrating and applying the no slip condition on the substrate $u|_{y=0}=0$, we find an expression for the velocity profile:
\[
u = -M^{1/n} \frac{n}{n+1} \left| \frac{\partial p}{\partial x} \right|^{1/n-1}\frac{\partial p}{\partial x} \left[h^{1/n+1} - (h-y)^{1/n+1}\right], \qquad 0 < y <h.
\]
Finally, integrating to find the flux,
\begin{equation}
q = -\mathcal M h^{2 + 1/n} \left| \frac{\partial p}{\partial x} \right|^{1/n-1}\frac{\partial p}{\partial x}, \qquad \mathcal M = \frac{n M^{1/n}}{(2n+1)}.
\label{eq:formflux}
\end{equation}
This equation holds for any choice of the applied pressure on the interface.  In the present study, we consider the competing effects of surface tension and van der Waals force, which is represented with a disjoining pressure term.  Again, under the lubrication approximation, these two effects result in
\begin{equation}
p = -\sigma\frac{\partial^2h}{\partial x^2} + \frac{A^*}{6\pi h^3}.
\label{eq:formpressure}
\end{equation}
Here $\sigma$ is the surface tension, and $A^*$ is the Hamaker constant.

In the absence of any external horizontal length scale (e.g. domain size), the system may be nondimensionalised to remove all parameters apart from the power-law exponent $n$.  Let the vertical length scale $[h]$ be equal to the mean thickness, and define a horizontal length scale $[x]$ and time scale $[t]$ by
\[
[x] = \sqrt{\frac{2\pi\sigma}{A^*}} [h]^2, \qquad [t] = \frac{[x]^{1+3/n}}{\mathcal M \sigma^{1/n}[h]^{1+2/n}}.
\]
Under this choice of nondimensionalisation, and substituting the expression for the pressure \eqref{eq:formpressure} into the flux \eqref{eq:formflux}, we arrive at the equations for the now dimensionless thickness $h(x,t)$ and flux $q(x,t)$:
\begin{equation}
\frac{\partial h}{\partial t} + \frac{\partial q}{\partial x} = 0, \qquad q = h^{2+1/n}\left|\frac{\partial^3 h}{\partial x^3} + \frac{1}{h^4}\frac{\partial h}{\partial x}\right|^{1/n-1}\left(\frac{\partial^3 h}{\partial x^3} + \frac{1}{h^4}\frac{\partial h}{\partial x}\right).
\label{eq:pde}
\end{equation}
This equation is of the form studied numerically in \cite{garg2017self-similar,hwang1993rupture}, as well as the limiting case (for large cylinder) of the cylindrical case considered in \cite{gorla2001rupture}.

\section{Numerical simulation of PDE}
\label{sec:numerics}

We simulate the thin film equation \eqref{eq:pde} using a finite-difference method, in which the flux $q$ is first computed from the values of thickness $h$ at each node using a five-point stencil for the first and third derivatives, and then a central finite difference is taken to find the evolution of thickness $h$.  This procedure avoids attempting to calculate the fourth spatial derivative of $h$, which may not exist everywhere for non-integer $n$ \cite{dallaston2024capillary}.  The solution is advanced in time using the implicit algorithm \texttt{ode15s} in MATLAB, with the banded (width-7) structure of the Jacobian matrix used to optimise the performance.

For completeness, boundary and initial conditions must be specified.  Our primary focus will be on local behaviour of solutions close to rupture, on which the boundary conditions will have little effect.  We choose periodic boundary conditions on a domain size $L$ which we will choose to be sufficiently large to see instability.  For initial conditions, as the system is nondimensionalised with respect to the average film thickness, we set
\[
h(x,t) = 1 + a\cos\left(\frac{2\pi x}{L}\right),
\]
where $a \ll 1$ is the magnitude of perturbation to a flat film, so that the mean nondimensional thickness is equal to unity.

\begin{figure}
\centering
\includegraphics{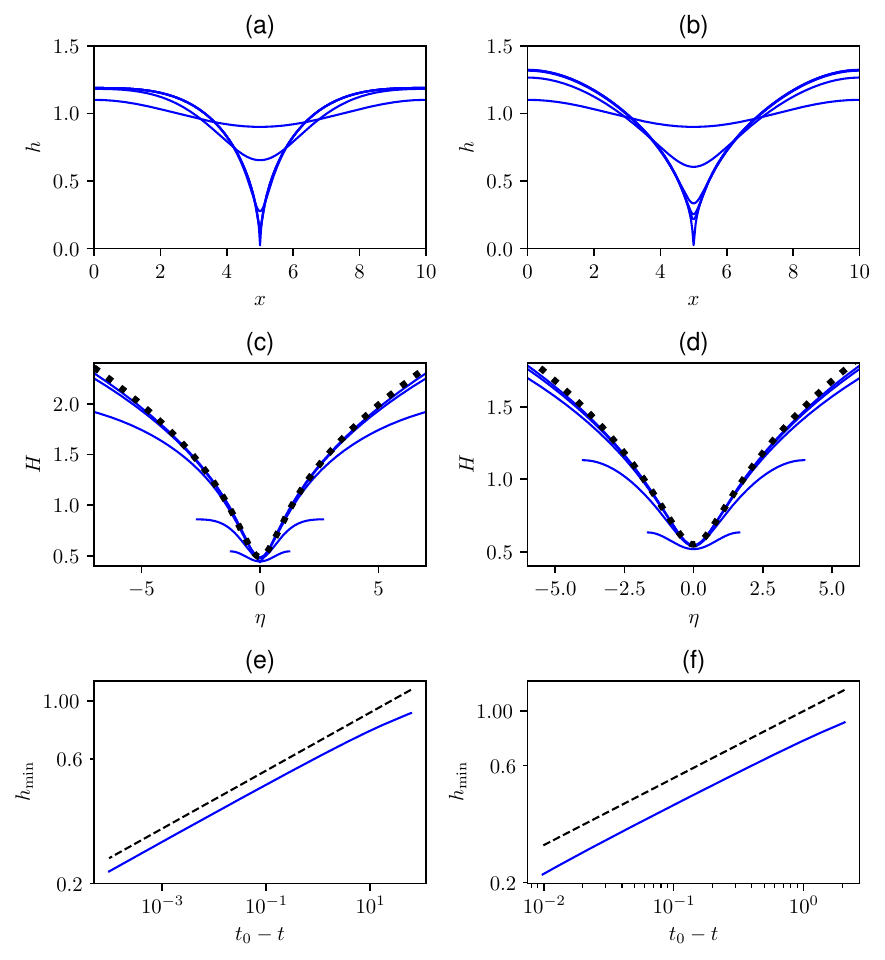}
\caption{Results of simulations of the PDE \eqref{eq:pde} for (a, c, e) the shear-thinning case $n=0.5$, and (b, d, f) the shear-thickening case.  In (a,b), the profiles $h(x,t)$ are plotted for times approaching rupture ($h\to 0$ at a point).  In (c,d), the profiles are scaled on to the similarity variables $\eta, H$ from \eqref{eq:simscalings}, and shown to asymptote to the numerically computed similarity solution \eqref{eq:simodes} (dotted line) described in Section \ref{sec:selfsim}.  In (c,d), the minimum thickness $h_\mathrm{min}(t)$ is plotted against time until rupture $t_0-t$.  The dashed line corresponds to the $n$-dependent exponent $\alpha = n/(n+4)$ predicted in \eqref{eq:simExponents}. }
\label{fig:pdeSimulations}
\end{figure}

In Fig.~\ref{fig:pdeSimulations}(a,b) we depict the results of numerical simulations of \eqref{eq:pde} for a typical shear-thinning and shear-thickening case: $n= 0.5$ and $n=1.5$, respectively.  These simulations each use $N=501$, a domain size $L=10$, and an initial perturbation of magnitude $a = 0.1$.  In each case, the initial depression in the centre of the domain is seen to deepen, and at a finite time the thickness becomes zero at a single point, while the velocity at this point becomes infinite.  The finite-time singularity of the PDE corresponds to the physical phenomenon of rupture.  By scaling the spatial coordinates $x$ and $h$ appropriately (as described in the next section), the solution is seen to asymptotically approach a universal profile, depicted in Fig.~\ref{fig:pdeSimulations}(c,d), respectively.  Finally, plotting the minimum thickness $h_\mathrm{min}(t) = \min_{x} h(x,t)$ as a function of estimated time until rupture, on a logarithmic scale, shows that the minimum thickness does indeed behave as a power law; this power law is that predicted in the self-similar analysis of the next section.  These results confirm that self-similar behaviour known to occur in the Newtonian case are also seen for power-law fluids, although both the scaling and the shape of similarity profiles are dependent on the power-law index $n$.

\section{Similarity solutions}
\label{sec:selfsim}

\subsection{Formulation}

The rupture process observed in the numerical simulations in the previous section suggests that the behaviour in the neighbourhood of the point of rupture will be self-similar.  In the Newtonian case, similarity solutions have previously been computed in \cite{witelski1999stability, zhang1999similarity}, which we generalise now for a non-Newtonian fluid of power-law type.  

Assume that rupture occurs at a location $x_0$ and time $t_0$, and define the similarity coordinate $\eta$ and scaled profile $H$ and flux $Q$ by
\begin{align}
h(x,t) &= \lambda^{1/2}(t_0-t)^{n/(n+4)} H(\eta), \\
\eta &= \frac{\lambda (x-x_0)}{(t_0-t)^{\beta}}, \\
 q(x,t) &= \lambda^{1-2/n}(t_0-t)^{\gamma}Q(\eta),
\label{eq:simscalings}
\end{align}
where the similarity exponents are
\begin{equation}
\alpha = \frac{n}{n+4}, \qquad \beta = \frac{2n}{n+4}, \qquad \gamma = \frac{2n-4}{n+4}.
\label{eq:simExponents}
\end{equation}
In these definitions, the similarity exponents have been chosen by balancing the time-dependence of each term in \eqref{eq:pde}, and match those found by \cite{garg2017self-similar}.  The value of $\alpha$ is also that seen in the decay of $h_\mathrm{min}$ in simulations in Fig.~\ref{fig:pdeSimulations}.  The scale factor $\lambda$ is defined to be
\[
\lambda = \left(\frac{n+4}{n}\right)^{2n/(n+4)},
\]
which simplifies the resulting system by removing an $n$-dependent prefactor.  After substituting these expressions into \eqref{eq:pde}, the equations for the scaled profile $H$ and flux $Q$ are
\begin{equation}
Q' = H - 2\eta H', \qquad Q^n = H^{1+2n}\left(H''' + \frac{H'}{H^4}\right),
\label{eq:simodes}
\end{equation}
where primes indicate ordinary derivatives with respect to $\eta$. 

Assuming symmetry in space, two boundary conditions at $\eta = 0$ are
\begin{equation}
H'(0) = Q(0) = 0.
\label{eq:simSymmetryBCs}
\end{equation}
For finite $n$, the condition $Q(0) = 0$ may be replaced with the condition $H'''(0) = 0$.  However, an expansion near $\eta=0$ reveals that for  $n < 1$, this third derivative will not be differentiable.  From the system \eqref{eq:simodes} and symmetry conditions \eqref{eq:simSymmetryBCs}, when $n < 1$ we have 
\begin{equation}
Q \sim H(0) \eta, \qquad H''' \sim \frac{Q^n}{H_0^{1+2n}} \sim H(0)^{-(1+n)} \eta^{n}, \qquad \eta \to 0^+.
\end{equation}
Integrating to find the series expansion for $H$:
\begin{equation}
H \sim H(0) + \frac{H''(0)}{2}\eta^2 + \frac{H(0)^{-(1+n)}}{(1+n)(2+n)(3+n)}\eta^{3+n}, \qquad \eta \to 0^+.
\label{eq:odeseriesexpansion}
\end{equation}
This phenomenon is related to the nonexistence of higher spatial derivatives of solutions to power-law lubrication flow at points where the pressure gradient vanishes, for instance as described in \cite{dallaston2024capillary}.  When it comes to numerical solution of the similarity profiles, particularly for small $n$,  more accuracy is obtained by formulating the problem in terms of the flux $Q$, which (unlike $H'''$) will be differentiable at the origin.

\subsection{Far-field condition}
\label{sec:wkbff}

In the far field we require the velocity in the original coordinates to be bounded, which implies
\begin{equation}
H - 2\eta H' \to 0, \qquad \eta \to \infty.
\label{eq:simFarfieldBCs}
\end{equation}
 
This constraint implies the leading-order behaviour of $H$ and $Q$ in the far field must be
\begin{equation}
H \sim c\eta^{1/2}, \qquad Q \sim d\eta^{1 - 2/n}, \qquad \eta \to \infty,
\label{eq:simFarfieldBehaviour}
\end{equation}
where $c$ is a positive constant, and 
\begin{equation}
d = c^{2+1/n}\left(\frac{3c}{8} + \frac{1}{2c^3}\right)^{1/n} > 0.
\label{eq:definitionOfd}
\end{equation}
Similar to the Newtonian case analysed by \cite{zhang1999similarity}, A WKB expansion around this far-field behaviour reveals that the condition  \eqref{eq:simFarfieldBCs} imposes two constraints on the far-field condition, as we now show.  Consider the exponential corrections to the far-field expansions for $H$ and $Q$:
\[
H \sim c\eta^{1/2} + \ldots + \hat H(\eta)\e^{k\eta^r}, \qquad Q \sim d\eta^{1-2/n} + \ldots + \hat Q(\eta) \e^{k\eta^r},
\]
where we suppress the higher-order algebraic terms in the expansions.  Here $k$ and $r$ are to be determined, with the signs of possible values of $k$ (or its real part) the important aspect.  The prefactor functions $\hat H$ and $\hat Q$ are not determined from the leading-order WKB analysis.  Substituting these expressions into \eqref{eq:simodes} and taking the leading order exponential terms, we find
\[
\hat Q = -2\eta \hat H, \qquad n\bar Q^{n-1}\hat Q = \bar H^{1+2n}\left(rk\eta^{r-1}\right)^3 \hat H.
\]
There are three different possible values for $k$, corresponding to three different WKB modes that exist near $H =c\eta^{1/2}$ in the far field.  By eliminating the functions $\tilde H$ and $\tilde Q$, and matching the coefficients and powers of $\eta$, we obtain
\[
r = \frac{2}{3n} + \frac{1}{6}, \qquad (rk)^3 = -\frac{2n d^{n-1}}{c^{2n+1}} < 0.
\]
There is thus one negative real $k$ and two complex values of $k$ with positive real part, that is,
\[
k = \frac{1}{r}\left(\frac{2nd^{n-1}}{c^{1+2n}}\right)^{1/3} \left\{-1, \e^{\mathrm i\pi/3}, \e^{-\mathrm i \pi/3}\right\}.
\]
As such, one mode decays as $\eta\to\infty$ and two modes grow in oscillatory fashion.  In order for the far field condition \eqref{eq:simFarfieldBCs} to be satisfied, both oscillatory modes must be identically zero, which imposes two constraints in the far field.  Along with the two symmetry conditions, \eqref{eq:simSymmetryBCs}, four boundary conditions are thus specified for the fourth-order boundary value problem \eqref{eq:simodes}.

\subsection{Numerical continuation}
\label{sec:continuation}
In order to solve the boundary-value problem \eqref{eq:simodes}, along with boundary conditions \eqref{eq:simFarfieldBCs} and \eqref{eq:simSymmetryBCs}, we use the numerical continuation package \textsc{Auto}07p \cite{doedel2007auto}.  In the context of thin-film similarity solutions, this method has been previously used in both the calculation of multiple solutions via the principle of homotopy continuation \cite{tseluiko2013homotopy}, as well as construction of solution branches over parameter space \cite{dallaston2021regular, dallaston2018discrete,dallaston2017self}.

\begin{figure}
\centering
\includegraphics{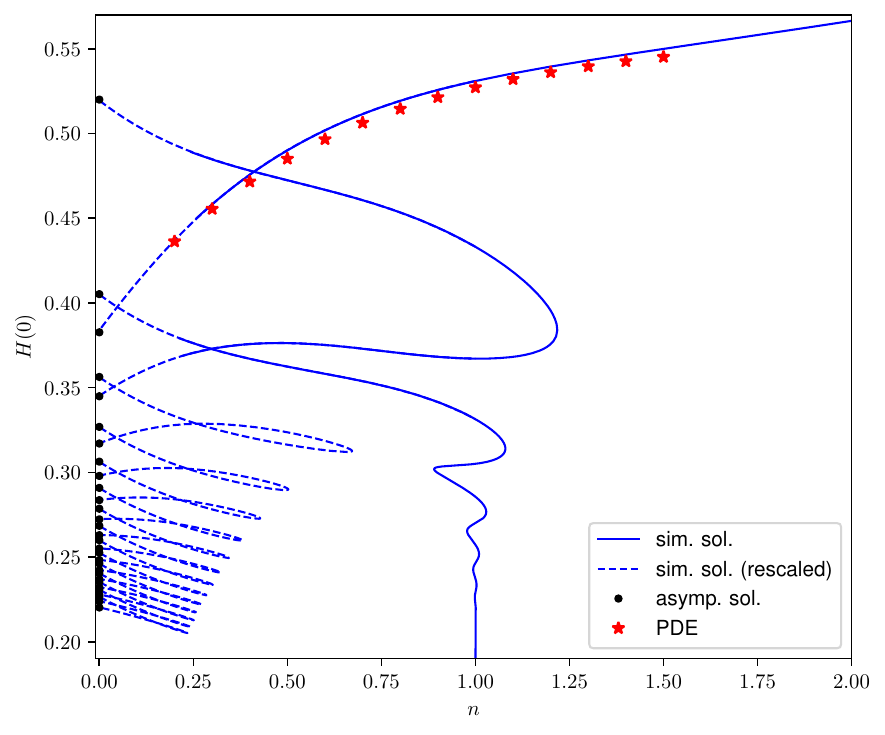}
\caption{Bifurcation diagram of similarity solutions (solutions of \eqref{eq:simodes}) for rupture of a power-law fluid depending on the power-law exponent $n$.  Solid lines are those computed using the continuation procedure described in subsection \ref{sec:continuation}, while the dashed lines are found using a rescaled method that converges for small $n$ described in subsection \ref{sec:rescaledContinuation}.  The dots indicate $n\to 0$ asymptotic solutions found in section \ref{sec:asymptotics}.  Stars indicate the results of numerical PDE simulations, strongly suggesting that only the primary branch is stable.}
\label{fig:branches}
\end{figure}

From \eqref{eq:simodes} we define a system of four ordinary differential equations with unknowns $H$, $H'$, $H''$, and $Q$  on the domain $0 < \eta < L$, with $L$ taken to be the far field:
\begin{equation}
\label{eq:simodesystem}
\frac{\mathrm d}{\mathrm d\eta} \begin{bmatrix} H \\ H' \\ H'' \\ Q \end{bmatrix} = \begin{bmatrix} H' \\ H'' \\ \displaystyle \frac{Q^n}{H^{1+2n}} - \frac{H'}{H^4} \\ H - 2\eta H' \end{bmatrix}, \qquad 0 < \eta < L,
\end{equation}
and boundary conditions \eqref{eq:simSymmetryBCs} at $\eta=0$, and far-field conditions implemented at $L \gg 1$:
\begin{equation}
H(L) - 2LH'(L) = 0, \qquad H'(L) + 2LH''(L) = 0. \label{eq:simNumericalFarfieldBCs}
\end{equation}
These two conditions in the far field enforce the far-field behaviour \eqref{eq:simFarfieldBehaviour} and represent the elimination of the two growing WKB modes detailed in subsection \ref{sec:wkbff}. In our computations we find that taking $L = 8$ results in sufficient accuracy.
In the Newtonian case ($n=1$), it is known that there are a countably infinite number of similarity solutions, only the first of which is linearly stable to perturbations.  Starting from these known solutions for the Newtonian case, we construct a bifurcation diagram by numerical continuation in the power-law exponent $n$.  We plot the results of this calculation in Fig.~\ref{fig:branches}, using the value $H(0)$ as a characteristic parameter representing each solution.  

The most notable features of the resulting bifurcation diagram shown in Fig.~\ref{fig:branches} are as follows.  Firstly, for $n > 1$, only the primary branch may be continued to arbitrarily large $n$ (the primary branch is the one that has largest $H(0)$ when $n=1$).   Although we do not report results for $n>2$, as we are mainly interested in the bifurcation structure for $n\lesssim 1$, our numerical observations suggest that this branch continues for arbitrarily large $n$.  All other branches annihilate via fold bifurcations with each other at points $n>1$, with the fold bifurcations approaching $n=1$ for higher branches (smaller $H(0)$).  For $n <1$, only the first four branches (the four largest $H(0)$ when $n=1$) continue towards $n\to 0$.  The other branches also annihilate via fold bifurcations, thereby resulting in a snaking bifurcation around $n=1$.  
We note that a number of points at which branches appear to cross over at certain values of $n$ are purely due to the use of $H(0)$ as a single representative value of a solution of a boundary-value problem, and these points do not correspond to bifurcations.

\begin{figure}
\centering
\includegraphics{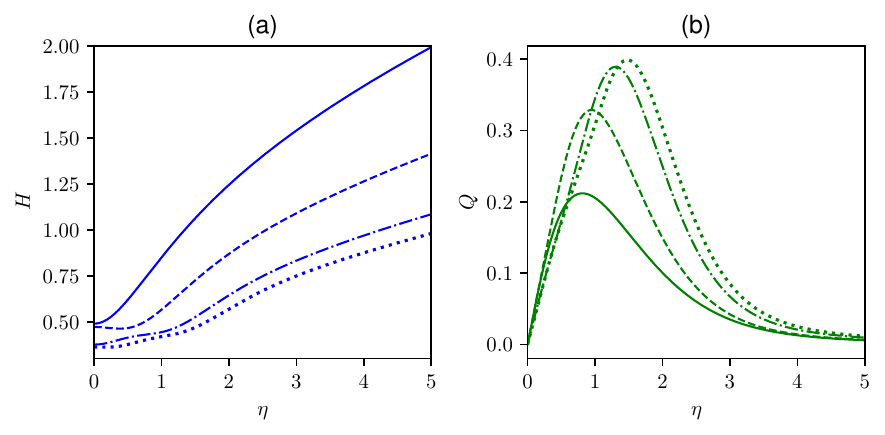}
\caption{The (a) similarity profiles $H$ and (b) fluxes $Q$ of the four similarity solutions that exist at $n=0.5$. The primary branch is denoted as solid, while the branches (in order of decreasing $H(0)$ are depicted as dashed, dash-dotted, and dotted lines, respectively.  Oscillations in the similarity profiles may be observed for this value of $n$.}
\label{fig:solsn05}
\end{figure}

To confirm which branch acts as an attractor for simulations of the PDE \eqref{eq:pde}, we estimate the prefactor in the minimum thickness in those simulations, which should be asymptotically related to $H(0)$ as
\[
h_\mathrm{min}(t) \sim (t_0-t)^\beta \lambda^{1/2} H(0), \qquad t \to t_0.
\]
In particular, $H(0)$ is estimated from the simulation for each $n$ by computing $h_\mathrm{min}(t)$ at the time $t$ at which $t_0-t = 0.05$, a time chosen to be sufficiently after the effect of the initial condition has decayed, but before numerical error has a significant effect.  The resulting values, shown as stars in Fig.~\ref{fig:branches}, are seen to correspond to the primary branch of solutions that continues to both $n\to 0$ and $n\to \infty$.  This branch is the continuation of the stable solution at $n=1$.  The similarity profiles on this branch at $n=0.5$ and $n=1.5$ are plotted in Figs.~\ref{fig:pdeSimulations}c and d, respectively, in order to show agreement with the numerical simulations at these values.

To observe the nature of solutions on higher branches, in Fig.~\ref{fig:solsn05} we plot the first four similarity profiles themselves (both $H$ and $Q$) at $n=0.5$.   The primary branch is also the one shown to agree with the numerical simulations in Fig.~\ref{fig:pdeSimulations}.  At this value of $n$, higher branches are seen to be oscillatory in $\eta$.   These oscillations have previously been observed in higher derivative terms in the Newtonian case \cite{witelski1999stability} where, in the high branch number (i.e. $H(0)\to 0$) limit,  these oscillations may be understood as arising from exponentially small terms that are present on part of the real line due to Stokes' phenomenon.  These oscillations play a role in the discrete selection of solutions as an appropriate wavelength is required for the boundary conditions to be satisfied \cite{chapman2013exponential}.  For smaller $n$, these oscillations are larger in amplitude and are apparent in the similarity profiles themselves.

Of the solution branches that may be found by numerically continuing from the Newtonian solutions at $n=1$, only four branches approach small values of $n$.  In the regime $n \lesssim 0.2$, the numerical method used to solve \eqref{eq:simodesystem} ceases to converge, indicating the need for a modified system to capture similarity solutions in this regime.  We describe such a scheme in the next subsection.  

\subsection{Rescaled solution for small $n$}
\label{sec:rescaledContinuation}
While the above scheme allows us to accurately compute similarity solutions when $n = O(1)$, the first four branches that approach $n=0$ cease to numerically converge when $n$ becomes small, in our case for $n \lesssim 0.2$.  The reason for this loss of numerical accuracy is due to $Q$ becoming exponentially small in an exponentially small inner region, which will become apparent when examining the small-$n$ asymptotic problem in section \ref{sec:asymptotics}.  Here, we introduce a rescaled version of the system \eqref{eq:simodes} to ameliorate this issue and allow us to compute solutions for small $n$.

Define a new independent variable $\zeta$ by
\[
\frac{\mathrm d \eta}{\mathrm d\zeta} = Q.
\]
Thus, for uniform $\zeta$, $\eta$ will change very slowly when $Q$ is small.  Under this transformation the system of equations \eqref{eq:simodes} becomes
\begin{equation}
\frac{\mathrm d}{\mathrm d\zeta} \begin{bmatrix} \eta \\ H \\ H' \\ H'' \\ Q \end{bmatrix} = Q \begin{bmatrix} 1 \\ H' \\ H'' \\ \displaystyle \frac{Q^n}{H^{1+2n}} - \frac{H'}{H^4} \\ H - 2\eta H' \end{bmatrix}, \qquad 0 < \zeta < \zeta_\infty,
\end{equation}
where we have introduced the additional equation for $\eta$ (note that primes in this equation still indicate $\eta$-derivatives).  The above transformation results in a system translationally invariant in $\zeta$, with the semi-infinite domain $\eta \in [0, \infty)$ mapped to the infinite domain $\zeta \in (-\infty, \infty)$.  To solve this system numerically we pose this system on the large domain $\zeta \in [0, \zeta_\infty]$ where $\zeta=0$ corresponds to a point $\eta = \delta \ll 1$ close to the origin, and $\zeta_\infty \gg 1$ represents the far field in $\eta$.  At $\eta = \delta$ we must use the series expansion \eqref{eq:odeseriesexpansion} to obtain the appropriate boundary conditions:
\begin{equation}
\begin{bmatrix}
\eta \\ H \\ H' \\ H'' \\ Q
\end{bmatrix} =
\begin{bmatrix}
\delta \\ H(0) \\ H''(0)\delta \\ \displaystyle H''(0) + \frac{H(0)^{-(n+1)}}{n+1} \delta^{n+1} \\ H(0)\delta
\end{bmatrix}, \qquad \zeta=0.
\label{eq:rescaledSimLeftBCs}
\end{equation}
Here $H(0)$ and $H''(0)$ are free parameters, so three boundary conditions (including the one for $\eta$) are applied at the left end of the domain.  The same far-field conditions as the unmodified problem are applied at $\zeta=\zeta_\infty$:
\begin{equation}
H - 2\eta H' = H' + 2\eta H''  = 0, \qquad \zeta = \zeta_\infty.
\end{equation}
There are thus five boundary conditions in total for the five variables.

\begin{figure}
\centering
\includegraphics{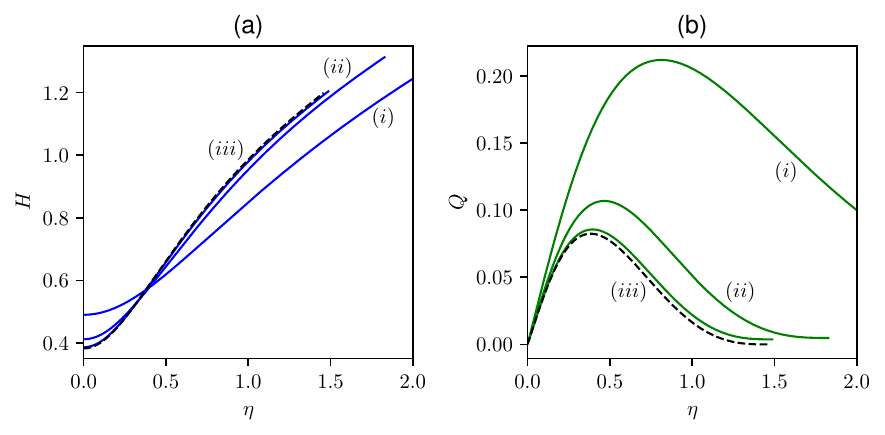}
\caption{The (a) similarity profiles $H$ and (b) flux $Q$ on the primary branch ($H(0)$ largest when $n=1$) for (i) $n=0.5$, (ii) $n = 0.1$, (iii) $n=0.01$.  The dashed line indicates the numerically computed leading-order asymptotic solution for $n\to 0$ found in section \ref{sec:asymptotics}.}
\label{fig:solsnearzero}
\end{figure}

We use this rescaled system to compute similarity solutions, again using the known Newtonian solution as a starting point, and decreasing $n$ for the first four branches.   In our computations we use numerical parameters $\delta = 0.01$, and $\eta_\infty = 100$, for which the solutions have converged.  For $n\gtrsim 0.2$, where both methods converge, the results from each are in agreement.  When $n \lesssim 0.2$, the rescaled method allows us to compute solutions for much smaller $n$, and $H(0)$ approaches finite values on each of these four branches as $n \to 0$.  In Fig.~\ref{fig:solsnearzero} we plot solutions on the primary branch for decreasing $n \in\{0.5, 0.1, 0.01\}$, demonstrating the convergence of the solution to an asymptotic limit as $n\to 0$.  

As well as the first four solution branches, we compute higher solution branches in the neighbourhood of $n=0$, which are not connected to the previously found solution branches near $n=1$.  To find these solutions we use a homotopy continuation method, adding an (unphysical) error term $\epsilon$ to the boundary condition in $H'(0)$ in \eqref{eq:rescaledSimLeftBCs} and allowing $H(0)$ to decrease.  Values of $H(0)$ at which $\epsilon=0$ are then also solutions that satisfy the original boundary conditions.  These solution branches are also plotted in Fig.~\ref{fig:branches}.  Branches in this region annihilate as pairs as $n$ increases while, as $n\to 0$, each branch approaches a solution with finite $H(0)$ as $n\to 0$.  In the next section, we will see these points correspond to leading-order solutions of the $n\to 0$ asymptotic limit.  While we have computed the first 26 solution branches beyond the four that connect to the rest of the bifurcation diagram, it is apparent that there will be a countably infinite number of such solutions, with the location of the fold bifurcations connecting successive branches tending to $n=0$ in the limit $H(0) \to 0$.

\section{Similarity solutions in the small $n$ limit}
\label{sec:asymptotics}

To support the above calculations of branches, and to provide motivation for the coordinate rescaling that we needed to compute solutions for small $n$ in subsection~\ref{sec:rescaledContinuation}, we examine the asymptotic behaviour of similarity solutions (solutions of the system \eqref{eq:simodes}) as $n$ becomes small.  We will see that the asymptotic structure involves an outer region, where $H$ and $Q$ are of order unity, an inner layer region of exponentially small size, in which $Q$ becomes exponentially small, corresponding to a rapid change in the third derivative of $H$, and a further innermost region, which can then be matched to the appropriate far field behaviour \eqref{eq:simFarfieldBehaviour}.  

This asymptotic structure is similar to that seen in a first-order ODE model in \cite{kember2000exponential}, in which the limit of a small exponent was also studied.  Our system will prove more difficult, however, as our leading-order problem remains a high-order nonlinear system that can only be solved numerically.   We will only carry out the asymptotic analysis to leading order.  The solution of the outer problem will be seen to  match well with the numerical solutions for $n>0$ as $n$ becomes small.  We also demonstrate how the matching between successive regions occurs, culminating with the matching between the innermost region and the farfield condition \eqref{eq:simFarfieldBehaviour}.

\subsection{Outer region}

Assume $n \ll 1$, and let $H \sim H_0$ and $Q \sim Q_0$ to leading order.  Since $Q^n \to 1$ as $n\to 0$, the system \eqref{eq:simodes} to leading order is
\begin{equation}
H_0''' = \frac{1}{H_0} - \frac{H_0'}{H_0^4}, \qquad Q_0' = H_0 - 2\eta H_0'.
\label{eq:simodesmalln}
\end{equation}
The first of these is a third-order nonlinear equation for the thickness $H_0$, while the second allows the flux $Q_0$ to be calculated.  From the symmetry conditions \eqref{eq:simSymmetryBCs} we require $H_0'(0)=Q_0(0) = 0$, while $H_0(0)$ and $H_0''(0)$ are yet to be determined.   The leading-order problem \eqref{eq:simodesmalln} implies $H_0'''(0) = 1/H_0(0)$, which is consistent with the small-$\eta$ expansion of the full problem \eqref{eq:odeseriesexpansion} in the small-$n$ limit.

For large $\eta$, the asymptotic behaviour of these leading-order solutions is (for some constant $a>0$)
\[
H_0 \sim a\eta^2, \quad Q_0 \sim -a\eta^3, \qquad \eta\to\infty,
\]
which violates the required far-field behaviour \eqref{eq:simFarfieldBCs}, and implies at some point $Q_0$ becomes negative.  There will thus be a point $\eta^* > 0$ where $Q_0$ vanishes, near which the asymptotic series breaks down.  Near this point the asymptotic expansion must exhibit an inner region in which the slope of $Q$ rapidly decays, so that $Q$ itself does not vanish.  The leading-order solutions will then only be valid on the interval $\eta \in [0, \eta^*]$.

As we will see in subsection \ref{sec:innerProblem}, the presence of the third derivative in \eqref{eq:simodes} means that in order to obtain a nontrivial solution in an inner region, $Q_0$ must vanish as a cubic as $\eta\to\eta^{*-}$; that is,
\begin{equation}
Q_0(\eta^*) = Q_0'(\eta^*) = Q_0''(\eta^*) = 0.
\label{eq:Q0goesCubic}
\end{equation}
These conditions apply two further boundary conditions on the outer problem, so that, along with the symmetry conditions, the correct number of boundary conditions are specified.  These conditions imply that, at $\eta = \eta^*$,
\begin{equation*}
H_0' = \frac{H_0}{2\eta^*}, \quad H_0'' = -\frac{H_0}{4\eta^{*2}}, \qquad \eta = \eta^*
\end{equation*}
and thus
\begin{equation*}
Q_0''' = \frac{3H_0}{4\eta^{*2}} - 2\eta^*\left(\frac{1}{H_0} - \frac{1}{2\eta^*H_0^3}\right), \qquad \eta = \eta^*.
\end{equation*}
As $\eta\to\eta^*$, we will see the profile $H(\eta)$ approaches its far-field behaviour, so define a constant $c$ such that
\begin{equation}
H_0 = c\eta^{*1/2}, \qquad \eta = \eta^*.
\label{eq:H0goesSquareRoot}
\end{equation}
In terms of the constant $c$ we have
\begin{equation}
Q_0''' = 2\eta^*\left[C - \frac{1}{c\eta^{*1/2}}\right], \qquad \eta = \eta^*,
\label{eq:cubicQ0Outer}
\end{equation}
where for convenience we define the constant
\[
C = \left(\frac{3c}{8} + \frac{1}{2c^3}\right)\eta^{*-5/2}.
\]
We will see the coefficient \eqref{eq:cubicQ0Outer} matches to a solution in an inner region near $\eta = \eta^*$ in subsection \ref{sec:innerProblem}.

\subsection{Numerical calculation of leading-order outer problem}
\label{sec:outerProblemComputation}
 Before carrying out the asymptotic analysis in the inner regions, we describe the numerical solution of the leading-order outer problem \eqref{eq:simodesmalln}, along with the symmetry conditions at $\eta=0$ and the conditions \eqref{eq:Q0goesCubic} at $\eta = \eta^*$.  This problem is still too intractable to solve exactly.  To compute numerical solutions, we again use \textsc{Auto}07p, with the boundary values $H(0), H''(0)$, and the domain size $\eta^*$ as unknown parameters.  To first obtain one solution of this equation from an arbitrary initial guess, we numerically continue in these three parameters, until we satisfy all of the conditions  \eqref{eq:Q0goesCubic}.  Then, to find other solutions of this system, we construct a homotopy continuation by allowing $H'(0)$ to be a free parameter.  By allowing $H'(0)$ to vary, we increase $\eta^*$ (while allowing other parameters to vary in order to keep the other boundary conditions satisfied), and for any value where $H'(0) = 0$, we obtain another solution of the leading-order problem.  

\begin{figure}
\centering
\includegraphics{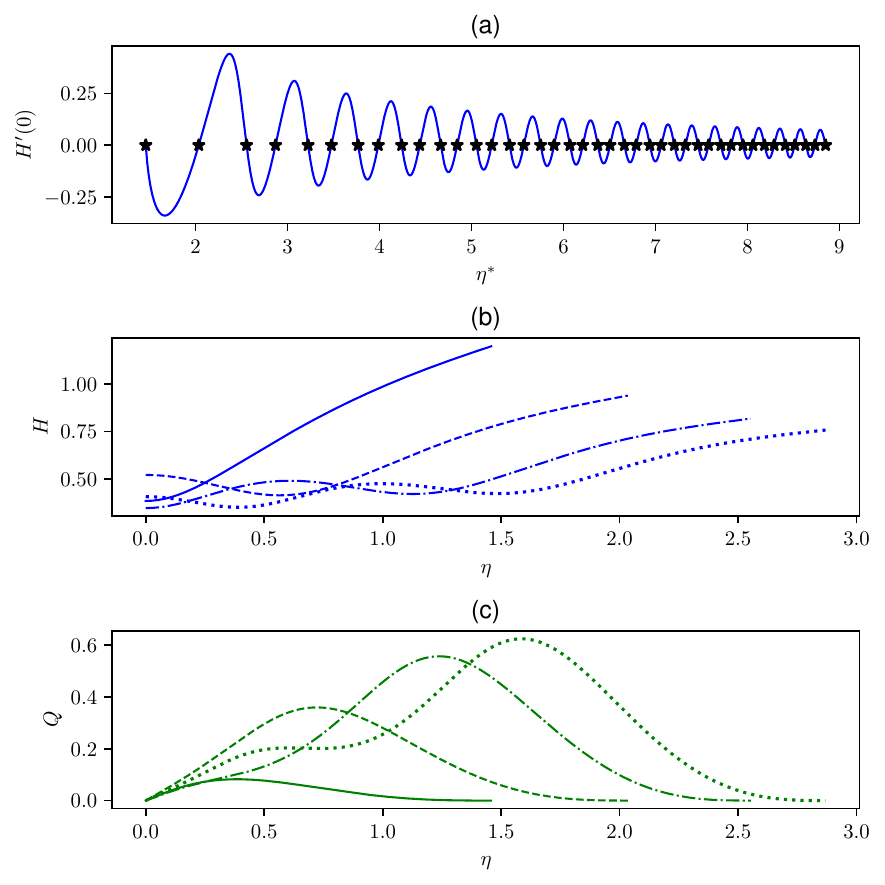}
\caption{Numerically calculated leading-order similarity profiles in the small-$n$ limit.  (a) the selection curve ($H'(0)$ as dependent on touchdown location $\eta^*$), that demonstrates a countably infinite number of solutions to \eqref{eq:simodesmalln} (corresponding to points where $H'(0)=0$).  (b) the first four similarity profiles $H_0$ (in order of increasing $\eta^*$), depicted as solid, dashed, dash-dotted, and dotted lines, respectively. (c) the corresponding scaled fluxes $Q_0$ for each of the first four similarity profiles.}
\label{fig:smallnProfiles}
\end{figure}

This selection curve ($\eta^*$ against $H'(0)$) and the first four leading-order profiles $H_0$ and $Q_0$, are depicted in Fig.~\ref{fig:smallnProfiles}.  Notably, the oscillatory behaviour seen in higher branches for $n=0.5$ in the full problem (Fig.~\ref{fig:solsn05}) is seen to a greater degree in the leading-order solutions to \eqref{eq:simodesmalln}.   The comparison between the leading-order profiles and the profiles of the full problem is depicted in Fig.~\ref{fig:solsnearzero}, demonstrating the validity of the leading order solution.  We have also included the computed values of $H_0(0)$ on the bifurcation diagram Fig.~\ref{fig:branches}, showing they are the limit points of the branches as $n\to 0$.

\subsection{Inner problem}
\label{sec:innerProblem}

We now consider the asymptotic behaviour of solutions near the point $\eta^*$ where the leading order-outer solution $Q_0(\eta^*) = 0$. Define a new independent variable $\xi$ by
\begin{equation}
\eta = \eta^* - \e^{-\xi/n},
\end{equation}   
where $\xi = O(1)$ in the inner region, and let
\begin{equation}
Q = \e^{-3\xi/n}\left(\tilde Q(\xi) + O(n)\right), \qquad H = c\eta^{1/2} + \e^{-3\xi/n}\left(\tilde H(\xi) + O(n)\right).
\end{equation}
Substituting into the system \eqref{eq:simodes} results in, to leading order in $n$:
\begin{equation}
\e^{-3\xi} = c\eta^{*1/2} \left(-6\tilde H + C\right), \qquad \tilde Q = -2\eta^*\tilde H.
\end{equation}
Thus
\begin{equation}
\tilde Q = \frac{\eta^*}{3}\left[\frac{\e^{-3\xi}}{c\eta^{*1/2}} - C \right].
\label{eq:innerQ}
\end{equation}
In terms of the outer variables $Q$ and $\eta$, the small-$n$ expansion of $\tilde Q$ is
\[
Q \sim \frac{\eta^*}{3}\left[C - \frac{\exp[-3n\log(\eta^* - \eta)]}{c\eta^{*1/2}}\right](\eta^* - \eta)^3 \sim \frac{\eta^*}{3}\left[C - \frac{1}{c\eta^{*1/2}}\right](\eta^*-\eta)^3, \qquad n \to 0,
\]
which matches the $\eta\to\eta^*$ behaviour of the outer solution \eqref{eq:cubicQ0Outer}.

From \eqref{eq:innerQ}, there is a point $\xi_0$ where $\tilde Q$ itself vanishes, namely,
\begin{equation}
\xi_0 = -\frac{1}{3}\log\left(cC\eta^{*1/2}\right).
\end{equation}
There will thus be a further, innermost region near this point.  As $\xi \to \xi_0$ we have
\begin{equation}
\tilde Q \sim C\eta^*(\xi_0-\xi), \qquad \xi\to\xi_0,
\label{eq:linearTildeQInner}
\end{equation}
which we will show matches to the innermost solution.

\subsection{Innermost region}

To scale into the innermost region where $\xi \to \xi_0$, we define the independent variable
\[
\eta = \eta^* + \frac{1}{n^{1/3}} \e^{-\xi_0/n} z,
\]
where $z = O(1)$ in the innermost region, and let
\[
Q = \e^{-3\xi_0/n}\left(q(z) + O(n)\right), \qquad H = c\eta^{*1/2} + \e^{-3\xi_0/n}\left(h(z) + O(n)\right).
\]
Substituting into the governing system \eqref{eq:simodes}, we have
\begin{equation}
\e^{-3\xi_0}\left(1 + n\log q(z)\right) = c\eta^{*1/2}\left(nh'''(z) + C\right) + nK + O(n^2), 
\label{eq:innermostProblem}
\end{equation}
along with $q'(z) = -2\eta^*h'(z) + O(n)$.  The constant $K = cC\eta^{*1/2}\log(c^2\eta^*)$ arises from the expansion of $H^{1+2n}$ for small $n$.

The $O(1)$ terms in \eqref{eq:innermostProblem} are identically equal, while the $O(n)$ terms, after eliminating $h$, give the following equation for $q$:
\begin{equation}
q'''(z) = -2C\eta^*\log\left(\frac{q(z)}{c^2\eta^*}\right).
\label{eq:innermostSolution}
\end{equation}
This equation is third-order and nonlinear, with no closed-form solution.  However, its far-field behaviour can be readily determined.  As $z\to -\infty$, $q$ becomes large and the third derivative and logarithmic term are in balance, resulting in
\begin{equation}
q \sim -C\eta^*|z|^3\log|z|, \qquad z \to -\infty.
\end{equation}
This farfield behaviour matches to the inner solution; writing the innermost solution for $z\to -\infty$ in the inner variable coordinates $\xi$ and $\tilde Q$, we have $q = \e^{3(\xi_0-\xi)/n}\tilde Q$, and $z = n^{1/3}\e^{(\xi_0-\xi)/n}$, so that
\[
\tilde Q \sim C\eta^* n \log\left(n^{1/3}\e^{(\xi_0-\xi)/n}\right) \sim C\eta^* (\xi_0-\xi), \qquad n \to 0,
\]
which matches the $\xi\to\xi_0$ behaviour of the inner solution \eqref{eq:linearTildeQInner}.

On the other hand, as $z\to\infty$, we require $q$ to approach a constant in order to match onto the farfield behaviour of the full problem \eqref{eq:simFarfieldBehaviour}.  Thus
\begin{equation}
\label{eq:ffq}
q \sim c^2\eta^*, \qquad z \to \infty.
\end{equation}
Linearising near this value, there is one decaying mode as well as two oscillatory growing modes that must be absent, similar to the farfield WKB analysis of the full problem (see subsection \ref{sec:wkbff}).

We now show the large-$z$ limit of the innermost problem matches to the farfield condition of the original problem in the small $n$ limit.  If we consider the far field behaviour of $Q(\eta)$ from \eqref{eq:simFarfieldBehaviour}, expressed in terms of $z$, and take $n\to 0$, we find
\[
Q \sim d\left(\eta^*+ n^{-1/3}\e^{-\xi_0/n} z\right)^{1-2/n} \sim c^2\eta^*\exp\left[\frac{\log(cC\eta^{*1/2})}{n}\right] = \e^{-3\xi_0/n} c^2\eta^*,
\]
where we have used the definition of the coefficient $d$ from \eqref{eq:definitionOfd}.  This term matches with the large-$z$ limit of $q$  \eqref{eq:ffq}.

\section{Discussion}
\label{sec:discussion}

Our major achievements in the above study are the calculation of similarity solutions for rupture of thin films of liquids with power-law rheology, and the extension of these calculations to the extreme-shear thinning regime, where an understanding of the asymptotic structure of solutions aids in developing an appropriate numerical scheme.  Two challenges that occur in our study are also likely to arise, not just in the calculation of similarity solutions, but across a wide range of interfacial problems involving power-law fluids.  Firstly, the nonexistence of higher derivatives of similarity profiles is a special case of the same phenomenon in thin film evolution of power-law fluids more generally (see e.g. \cite{dallaston2024capillary}).  This challenge is not major as long as numerical methods do not rely on calculation of these higher derivatives; we have found in both our PDE simulations and similarity solutions that using the flux as a primary variable avoids this issue.  

The more severe challenge occurs for small $n$, where the solution profile needs to be resolved in an exponentially small (in the limit $n\to 0$) region.  While in this work we have analysed the presence of such an exponentially small region in similarity solutions only, it is likely to be representative of the numerical difficulties of performing any kind of simulations of extreme shear-thinning ($n\ll 1$) fluids.  In the literature to date, most examples of simulations of interfacial power-law flow use values of $n$ relatively close to 1, most likely due to these numerical challenges.  Our analysis may also highlight a way forward, as some kind of dynamic mesh refinement could be used to concentrate numerical resolution in regions where the flux becomes very small.  Lending weight to this argument, the relatively small amount of literature that considers asymptotic analysis of extreme shear-thinning rheology in Stokes \cite{brewster1995asymptotics,chapman1997extrusion} or Hele--Shaw flow \cite{richardson2007saffman,richardson2015hele} tend to result in complicated boundary layer structures, sometimes with exponentially small regions.

A number of extensions particular to self-similar rupture of power-law fluids are also apparent.
We have not computed axisymmetric similarity solutions, which would be relevant for point rupture, or considered the stability of plane- or axisymmetric similarity solutions to determine which would be the general behaviour seen in two (horizontal) dimensional flow.  Results from the Newtonian case \cite{witelski2000dynamics} suggest that the axisymmetric case is more likely to be the stable one, but, as shown in \cite{dallaston2021regular}, the stability can depend on values of exponents in generalised thin film models, and so in the present model, could potentially depend on the power-law exponent $n$.

In addition, in applying the lubrication theory in our model we have tacitly assumed that viscosity, surface tension, and van der Waals forces are in dominant balance in the limit as rupture is approached, with inertia negligible.  If inertia is not neglected in the first instance, such as in the two-dimensional simulations of the full Navier-Stokes equations performed in \cite{garg2017self-similar}, or in the free-sheet simulations in \cite{thete2015self}, different flow regimes may be observed, dependent on both the time to rupture, as well as the value of the power-law exponent $n$.  Inertia notwithstanding, there may also be a crossover from lubrication theory to full two-dimensional flow where both $x$ and $y$ coordinates scale the same way (see \cite{moreno-boza2020stokes} for the Newtonian case, for example).  For the most part, in each of these regimes only the scalings are known at present, and the calculation of similarity profiles remains unresolved.  The methods and challenges described in this paper are likely to be applicable to computing similarity solutions in these regimes also.

\section*{Acknowledgments}
MCD and SWM acknowledge support from the Australian Research Council via the Discovery Project DP250101095.  SAK acknowledges additional funding from the Faculty of Science at QUT to support this research.


\section*{References}
\begin{thebibliography}{10}

\bibitem{barenblatt1996scaling}
G.I. Barenblatt.
\newblock {\em Scaling, self-similarity, and intermediate asymptotics:
  dimensional analysis and intermediate asymptotics}.
\newblock Cambridge University Press, 1996.

\bibitem{becker2003complex}
J.~Becker, G.~Gr{\"u}n, R.~Seemann, H.~Mantz, K.~Jacobs, K.R. Mecke, and
  R.~Blossey.
\newblock Complex dewetting scenarios captured by thin-film models.
\newblock {\em Nature materials}, 2:59--63, 2003.

\bibitem{bonn2009wetting}
D.~Bonn, J.~Eggers, J.~Indekeu, J.~Meunier, and E.~Rolley.
\newblock Wetting and spreading.
\newblock {\em Rev. Modern Phys.}, 81:739--805, 2009.

\bibitem{brewster1995asymptotics}
M.E. Brewster, S.J. Chapman, A.D. Fitt, and C.P. Please.
\newblock Asymptotics of slow flow of very small exponent power-law
  shear-thinning fluids in a wedge.
\newblock {\em Eur. J. Appl. Math.}, 6:559--571, 1995.

\bibitem{chapman2023role}
S.J. Chapman, M.C. Dallaston, S.~Kalliadasis, P.H. Trinh, and T.P. Witelski.
\newblock The role of exponential asymptotics and complex singularities in
  self-similarity, transitions, and branch merging of nonlinear dynamics.
\newblock {\em Physica D}, 453:133802, 2023.

\bibitem{chapman1997extrusion}
S.J. Chapman, A.D. Fitt, and C.P. Please.
\newblock Extrusion of power-law shear-thinning fluids with small exponent.
\newblock {\em Int. J. Nonlin. Mech.}, 32:187--199, 1997.

\bibitem{chapman2013exponential}
S.J. Chapman, P.H. Trinh, and T.P. Witelski.
\newblock Exponential asymptotics for thin film rupture.
\newblock {\em SIAM J. Appl. Math.}, 73:232--253, 2013.

\bibitem{craster2009dynamics}
R.V. Craster and O.K. Matar.
\newblock Dynamics and stability of thin liquid films.
\newblock {\em Rev. Modern Phys.}, 81:1131, 2009.

\bibitem{dallaston2024capillary}
M.C. Dallaston.
\newblock Capillary levelling of thin liquid films of power-law rheology.
\newblock {\em ANZIAM J.}, 66:62--76, 2024.

\bibitem{dallaston2021regular}
M.C. Dallaston, M.A. Fontelos, M.A. Herrada, J.M. Lopez-Herrera, and J.~Eggers.
\newblock Regular and complex singularities of the generalized thin film
  equation in two dimensions.
\newblock {\em J. Fluid Mech.}, 917:A20, 2021.

\bibitem{dallaston2018discrete}
M.C. Dallaston, M.A. Fontelos, D.~Tseluiko, and S.~Kalliadasis.
\newblock Discrete self-similarity in interfacial hydrodynamics and the
  formation of iterated structures.
\newblock {\em Phys. Rev. Lett.}, 120:034505, 2018.

\bibitem{dallaston2017self}
M.C. Dallaston, D.~Tseluiko, Z.~Zheng, M.A. Fontelos, and S.~Kalliadasis.
\newblock Self-similar finite-time singularity formation in degenerate
  parabolic equations arising in thin-film flows.
\newblock {\em Nonlinearity}, 30:2647, 2017.

\bibitem{doedel2007auto}
E.J. Doedel, A.R. Champneys, F.~Dercole, T.F. Fairgrieve, Yu.A. Kuznetsov,
  B.~Oldeman, R.C. Paffenroth, B~Sandstede, XJ~Wang, and CH~Zhang.
\newblock Auto-07p.
\newblock {\em Montreal Concordia University}, 2007.
\newblock {http://indy.cs.concordia.ca/auto/}.

\bibitem{eggers2015singularities}
J.~Eggers and M.A. Fontelos.
\newblock {\em Singularities: formation, structure, and propagation},
  volume~53.
\newblock Cambridge University Press, 2015.

\bibitem{eley2005applied}
R.~R. Eley.
\newblock Applied rheology in the protective and decorative coatings industry.
\newblock {\em Rheology Reviews}, 2005:173--240, 2005.

\bibitem{erneux1993nonlinear}
T.~Erneux and S.H. Davis.
\newblock Nonlinear rupture of free films.
\newblock {\em Phys Fluids A}, 5:1117--1122, 1993.

\bibitem{flitton2004surface}
J.C. Flitton and J.R. King.
\newblock Surface-tension-driven dewetting of {N}ewtonian and power-law fluids.
\newblock {\em J. Eng Math.}, 50:241--266, 2004.

\bibitem{fowler1996ice}
A.C. Fowler and C.~Johnson.
\newblock Ice-sheet surging and ice-stream formation.
\newblock {\em Ann. Glaciol.}, 23, 1996.

\bibitem{fowler2000small}
A.C. Fowler, G.~Kember, and S.G.B. O'Brien.
\newblock Small exponent asymptotics.
\newblock {\em IMA J. Appl. Math.}, 64:23--38, 2000.

\bibitem{garg2017self-similar}
V.~Garg, P.M. Kamat, C.R. Anthony, S.S. Thete, and O.A. Basaran.
\newblock Self-similar rupture of thin films of power-law fluids on a
  substrate.
\newblock {\em J. Fluid Mech.}, 826:455--483, 2017.

\bibitem{garg2022local}
V.~Garg, S.S. Thete, C.R. Anthony, and O.A. Basaran.
\newblock Local dynamics during thinning and rupture of liquid sheets of
  power-law fluids.
\newblock {\em J. Fluid Mech.}, 942:A15, 2022.

\bibitem{gaver1990dynamics}
D.P. Gaver and J.B. Grotberg.
\newblock The dynamics of a localized surfactant on a thin film.
\newblock {\em J. Fluid Mech.}, 213:127--148, 1990.

\bibitem{ghatak1999dynamics}
A.~Ghatak, R.~Khanna, and A.~Sharma.
\newblock Dynamics and morphology of holes in dewetting of thin films.
\newblock {\em J. Colloid Interface Sci.}, 212:483--494, 1999.

\bibitem{glasner2003coarsening}
K.B. Glasner and T.P. Witelski.
\newblock Coarsening dynamics of dewetting films.
\newblock {\em Phys. Rev. E}, 67:016302, 2003.

\bibitem{gorla2001rupture}
R.S.R. Gorla.
\newblock Rupture of thin power-law liquid film on a cylinder.
\newblock {\em J. Appl. Mech.}, 68(2):294--297, 2001.

\bibitem{huppert1982propagation}
H.E. Huppert.
\newblock The propagation of two-dimensional and axisymmetric viscous gravity
  currents over a rigid horizontal surface.
\newblock {\em J. Fluid Mech.}, 121:43--58, 1982.

\bibitem{huppert2006gravity}
H.E. Huppert.
\newblock Gravity currents: a personal perspective.
\newblock {\em J. Fluid Mech.}, 554:299--322, 2006.

\bibitem{hwang1993rupture}
C-C. Hwang and S-H. Chang.
\newblock Rupture theory of thin power-law liquid film.
\newblock {\em J. Appl. Phys.}, 74:2965--2967, 1993.

\bibitem{ida1996thin}
M.P. Ida and M.J. Miksis.
\newblock Thin film rupture.
\newblock {\em Appl. Math. Lett.}, 9:35--40, 1996.

\bibitem{iyer1996leveling}
R.R. Iyer and D.W. Bousfield.
\newblock The leveling of coating defects with shear thinning rheology.
\newblock {\em Chem. Eng. Sci.}, 51:4611--4617, 1996.

\bibitem{jensen1992insoluble}
O.E. Jensen and J.B. Grotberg.
\newblock Insoluble surfactant spreading on a thin viscous film: shock
  evolution and film rupture.
\newblock {\em J. Fluid Mech.}, 240:259--288, 1992.

\bibitem{kedda2025mathematical}
S.A. Kedda.
\newblock {\em Mathematical Models of Self-similarity in Thin Film Evolution}.
\newblock PhD thesis, Queensland University of Technology, 2025.

\bibitem{kedda2024long}
S.A. Kedda, M.C. Dallaston, and S.W. McCue.
\newblock Long-time emergent dynamics of liquid films undergoing
  thermocapillary instability.
\newblock {\em Phys. Rev. E}, 110:035104, 2024.

\bibitem{kember2000exponential}
G.C. Kember, A.C. Fowler, J.D. Evans, and S.B.G. O'Brien.
\newblock Exponential asymptotics with a small exponent.
\newblock {\em Quart. Appl. Mech.}, 58:561--576, 2000.

\bibitem{king2001two}
J.R. King.
\newblock Two generalisations of the thin film equation.
\newblock {\em Math. Comput. Model.}, 34:737--756, 2001.

\bibitem{miladinova2004thin}
S~Miladinova, G~Lebon, and E~Toshev.
\newblock Thin-film flow of a power-law liquid falling down an inclined plate.
\newblock {\em J. Non-Newton. Fluid}, 122:69--78, 2004.

\bibitem{moreno-boza2020stokes}
D.~Moreno-Boza, A.~Mart{\'\i}nez-Calvo, and A.~Sevilla.
\newblock Stokes theory of thin-film rupture.
\newblock {\em Phys. Rev. Fluids}, 5:014002, 2020.

\bibitem{myers1998thin}
T.~G. Myers.
\newblock Thin films with high surface tension.
\newblock {\em Siam Rev.}, 40:441--462, 1998.

\bibitem{myers2005application}
T.~G. Myers.
\newblock Application of non-{N}ewtonian models to thin film flow.
\newblock {\em Physical Review E}, 72:066302, 2005.

\bibitem{noble2013thin}
P.~Noble and J-P. Vila.
\newblock Thin power-law film flow down an inclined plane: consistent
  shallow-water models and stability under large-scale perturbations.
\newblock {\em J. Fluid Mech.}, 735:29--60, 2013.

\bibitem{obrien1998asymptotics}
S.B.G. O'Brien, E.G. Gath, A.C. Fowler, and G.~Kember.
\newblock Asymptotics with small exponent in a model for ice-sheet surging.
\newblock In {\em Math. Proc. R. Irish Acad.}, pages 67--80, 1998.

\bibitem{oron1997long-scale}
A.~Oron, S.H. Davis, and S.G. Bankoff.
\newblock Long-scale evolution of thin liquid films.
\newblock {\em Rev. Mod. Phys.}, 69:931--980, 1997.

\bibitem{richardson2007saffman}
G.~Richardson and J.R. King.
\newblock The {S}affman--{T}aylor problem for an extremely shear-thinning
  fluid.
\newblock {\em Quart. J. Mech. Appl. Math.}, 60:161--200, 2007.

\bibitem{richardson2015hele}
G.~Richardson and JR~King.
\newblock The {H}ele--{S}haw injection problem for an extremely shear-thinning
  fluid.
\newblock {\em Eur. J. Appl. Math.}, 26:563--594, 2015.

\bibitem{ruckenstein2018spontaneous}
E.~Ruckenstein and R.K. Jain.
\newblock Spontaneous rupture of thin liquid films.
\newblock In {\em Wetting Theory}, pages 588--603. CRC press, 2018.

\bibitem{shklyaev2012long}
S.~Shklyaev, A.A. Alabuzhev, and M.~Khenner.
\newblock Long-wave {M}arangoni convection in a thin film heated from below.
\newblock {\em Phys. Rev. E}, 85:016328, 2012.

\bibitem{shklyaev2010superexponential}
S.~Shklyaev, A.V. Straube, and A.~Pikovsky.
\newblock Superexponential droplet fractalization as a hierarchical formation
  of dissipative compactons.
\newblock {\em Phys. Rev. E.}, 82:020601, 2010.

\bibitem{teletzke1987liquids}
G.F. Teletzke, D.H. Ted, and L.E. Scriven.
\newblock How liquids spread on solids.
\newblock {\em Chem. Eng. Comm.}, 55:41--82, 1987.

\bibitem{thete2015self}
S.S. Thete, C.~Anthony, O.A. Basaran, and P.~Doshi.
\newblock Self-similar rupture of thin free films of power-law fluids.
\newblock {\em Phys. Rev. E}, 92:023014, 2015.

\bibitem{tseluiko2013homotopy}
D.~Tseluiko, J.~Baxter, and U.~Thiele.
\newblock A homotopy continuation approach for analysing finite-time
  singularities in thin liquid films.
\newblock {\em IMA J. Appl. Math.}, 78:762--776, 2013.

\bibitem{witelski1999stability}
T.P. Witelski and A.J. Bernoff.
\newblock Stability of self-similar solutions for van der {W}aals driven thin
  film rupture.
\newblock {\em Phys. Fluids}, 11:2443--2445, 1999.

\bibitem{witelski2000dynamics}
T.P. Witelski and A.J. Bernoff.
\newblock Dynamics of three-dimensional thin film rupture.
\newblock {\em Physica D}, 147:155--176, 2000.

\bibitem{zhang1999similarity}
W.W. Zhang and J.R. Lister.
\newblock Similarity solutions for van der {W}aals rupture of a thin film on a
  solid substrate.
\newblock {\em Phys. Fluids}, 11:2454--2462, 1999.

\bibitem{zhang2003analysis}
Y.L. Zhang, O.K. Matar, and R.V. Craster.
\newblock Analysis of tear film rupture: effect of non-{N}ewtonian rheology.
\newblock {\em J. Colloid Interface Sci.}, 262:130--148, 2003.

\end{thebibliography}
\end{document}